\documentstyle[epsfig,floatfig]{hep99}
\newcommand{\jp}    { J^P }
\newcommand{\dst}   {\mbox{$D^{\star \star}$}}
\newcommand{\dstz}  {\mbox{$D^{\star}_0$}}
\newcommand{\done}  {\mbox{$D_1$}}
\newcommand{\dstt}  {\mbox{$D^{\star}_2$}}
\newcommand{\dsst}  {\mbox{$D_s^{\star \star}$}}
\newcommand{\dpri}  {\mbox{$D^{\prime}$}}
\newcommand{\dstp}  {\mbox{$D^{\star \prime}$}}
\newcommand{\epem}  {\mbox{$e^+ e^-$}}
\begin{document}

\title{Production of excited charmed mesons \\ at LEP}

\author{D. Abbaneo}

\address{CERN, CH--1211, Geneva 23, Switzerland \\[3pt]
E-mail: {\tt Duccio.Abbaneo@cern.ch} }

\abstract{Studies of the production of orbitally excited charmed and 
charmed  strange mesons in \epem collisions, performed by the LEP
collaborations are reviewed. Measurements of the production rates of
orbitally excited charmed mesons in semileptonic $b$ decays are presented.
Searches for charmed meson radial excitations are also briefly discussed.
} 

\maketitle

\section{Introduction\label{sec:intro}}
In the hadronization of charm quarks into charmed mesons,
states with angular momentum $L=1$ can be produced, generically
indicated with \mbox{$D_{(s)}^{\star \star}$}, where the subscript
applies to charmed  strange mesons.

Four such states of the $c \bar q$ system are predicted ($q= u, d, s$),
one of total spin 0 (usually indicated as \dstz, $\jp = 0^+$ ), 
two of spin 1 (\done, $\jp=1^+$)
and one of spin 2 (\dstt, $\jp=2^+$). 
These excited states decay via strong interactions 
to $D$ or 
\mbox{$D^{\star}$} accompanied by a pion (a kaon in the case of the 
\dsst\ states). Spin--parity conservation implies that the \dstz\ decays
to $D \pi$ only, the \done\ to \mbox{$D^{\star} \pi$} only, while the 
\dstt\ can access both final states.

In the framework of HQET, for $m_Q \to \infty$, the light quark degrees 
of freedom decouple from the heavy quark, and new symmetries arise. In this
limit the \dstz\ and one of the two \done\ mesons are expected to decay in S 
wave only, and to be broad ($\Gamma$ around 100~MeV), the other \done\ and the
\dstt\ in D wave only, with narrow widths.
No clear evidence for the broad states has been found so far, while the two 
narrow states have been measured for all of the three systems 
\mbox{$D^{\star \star 0}$},
\mbox{$D^{\star \star +}$},
\mbox{$D^{\star \star}_s$}~\cite{PDG}.
At LEP \dst\ states can be produced in the fragmentation of charm quarks 
from $Z$ decays, or in the decay of bottom hadrons. 

The production rate of orbitally excited charmed mesons in semileptonic
decays of bottom mesons is a major $b$--physics issue. The exclusive
decay rates into \mbox{$D\ell \nu$} and \mbox{$D^{\star} \ell \nu$}
only account for 70--80\% of the total inclusive semileptonic rate.
The remaining part should be ascribed to narrow or wide \dst,
or non resonant four--body decays. The rates and the nature of final states
other than  \mbox{$D\ell \nu$} and \mbox{$D^{\star} \ell \nu$} is one
of the major sources of systematic uncertainty for many  $b$--physics 
measurements which make use of semileptonic final states such as
$\tau_{B^0_d}$, $\Delta m_d$ or $V_{cb}$.
Theoretical predictions exist on the ratio of \dstt\ to \done\ expected
in semileptonic b decays.

Radially excited charmed mesons are also predicted to exist, in two states,
the \dpri\ ($\jp=0^-$), and the \dstp\ ($\jp=1^-$), with dominant
decay modes \mbox{$D \pi \pi$} and \mbox{$D^{\star} \pi \pi$}, respectively.
While the first state is too difficult to reconstruct in present experiments,
the  \dstp\ has been searched for at LEP, with contradictory results.

\boldmath
\section{Production of \dst\ mesons in $Z$ decays}
\unboldmath
\subsection{The channels}

The cleanest channel to reconstruct \dst\ states is the 
{\mbox{$D^{\star +} \pi^-$}} final state, which takes advantage of the clear
experimental signature of the small mass difference between
the $D^{\star +}$ and the $D^0$ in the  {\mbox{$D^{\star +} \to D^0 \pi^+$}}
decay.
This channel has been used by ALEPH~\cite{A_d}, DELPHI~\cite{D_d}
and OPAL~\cite{O_d}, where OPAL reconstructs the  $D^0$ via 
{\mbox{$D^0 \to K^- \pi^+$}} decays, DELPHI uses also 
{\mbox{$D^0 \to K^- \pi^+ \pi^- \pi^+$}} and ALEPH the previous two plus
{\mbox{$D^0 \to K^- \pi^+ \pi^0$}}. 
The {\mbox{$D^{\star +} \pi^-$}} final state is accessible for both the
{\mbox{$D^0_1$}} and the 
{\mbox{$D^{\star 0}_2$}} mesons, and therefore the reconstructed candidates 
can be either of the two states.

\begin{table*}
\begin{center}
\caption{Results on \dst\ production probabilities in $Z$ hadronic decays.
\label{tab:dstst}
}
\begin{tabular}{cccc} 
\br
& ALEPH & DELPHI & OPAL \\
& (preliminary) & (preliminary) & (published)\\
\mr
$f(c\to D_1)$ 
& $0.032 \pm 0.009$ 
& $0.019 \pm 0.004$ & $0.021 \pm 0.008$ \\
$f(c\to D^{\star}_2)$ 
& $0.094 \pm 0.019$ 
& $0.047 \pm 0.013$ & $0.052 \pm 0.026$ \\
$f(b\to D_1)$ 
& $0.046 \pm 0.014$ 
& $0.020 \pm 0.006$ & $0.050 \pm 0.015$ \\
$f(b\to D^{\star}_2)$ 
& $<0.039\ @\ 95\% \ {\mathrm{CL}}$ 
& $0.048 \pm 0.020$ & $0.047 \pm 0.027$ \\
\br
\end{tabular}
\end{center}
\end{table*}

ALEPH has also studied the  {\mbox{$D^{+} \pi^-$}} final state, which has
a larger background but is accessible only by the  {\mbox{$D^{\star 0}_2$}}
meson. In this case the tighter kinematic cuts applied make the analysis 
sensitive to the charm contribution only (\dst\ from charm fragmentation 
have on average higher momentum than those from $b$ decays).

Finally ALEPH has also studied the channel {\mbox{$D^{0} \pi^+$}}, which is
a final state for  {\mbox{$D^{\star +}_2$}} decays. the $D^{0}$ is
reconstructed in the channels 
{\mbox{$D^0 \to K^- \pi^+$}} and 
{\mbox{$D^0 \to K^- \pi^+ \pi^- \pi^+$}}, and in this second, again,
kinematic cuts imposed to reduce the background damp the contribution of
$b$ events.

\subsection{Fitting methods and results \label{sec:boh}}

Samples enriched in $b$ and $c$ events are prepared, by applying lifetime
$b$--tagging in the hemisphere opposite to the reconstructed candidate,
combined with the information of the measured decay length 
(longer for $b$ events)
and the momentum (larger for $c$ events) of the reconstructed candidate.

ALEPH performs a simple fit to the  {\mbox{$D^{\star \star} - D^{\star}$}}
or {\mbox{$D^{\star \star} - D$}} mass difference, using a double
Breit--Wigner when both the \done\ and the \dstt\ are expected to contribute.

DELPHI and OPAL, that did not study the channels where only the \dstt\
contributes, use the helicity angle to help disentangling the 
two states. This is defined in the $D^{\star}$ rest frame
as the angle $\Theta_H$ between the $\pi$ from the $D^{\star}$ decay and the  $\pi$
from the \dst\ decay. It is expected to be distributed, in the HQET limit,
as  $1+3 \cos^2 \Theta_H$ in the case of \done\ decays and as $\sin^2 \Theta_H$
in the case of \dstt\ decays.

The fit yields the probability of observing the selected final state in a 
hadronic Z decay. Folding $Z$ partial widths, decay branching ratios
and selection efficiencies, the probability of producing a \dst\ state
from the fragmentation of a charm quark or the decay of a bottom hadron
can be derived, obtaining the results reported in Table~\ref{tab:dstst}.

Although no compelling discrepancy can be claimed, several differencies,
at the level of twice the estimated errors, are observed. In particular
ALEPH reports a significantly higher rate of \dstt\ from $c$ hadronization
than the two other experiments, DELPHI has a lower \done\ rate from $b$ decays,
and ALEPH finds no \dstt\ production in $b$ decays while the other 
collaborations find a signal of about $2\sigma$ significance. 
ALEPH and DELPHI results are still preliminary.

\boldmath
\section{\dst\ mesons in semileptonic $b$ decays}
\unboldmath

Two strategies have been adopted to study \dst\ production
and generic four--body final states in semileptonic $b$ decays.
In both cases the first step is to select 
{\mbox{$(D^+ \ell^-)$}}, {\mbox{$(D^0 \ell^-)$}}, 
{\mbox{$(D^{\star +} \ell^-)$}} pairs. Then one possibility
is to look for a charged pion that forms a heavier resonance with the 
$D^{(\star)}$ candidate, and extract the individual 
{\mbox{$B\to D^{(\star)}_J$}} rates. The other strategy is to 
search for pions incompatible with the primary vertex and 
compatible with the reconstructed $b$ vertex, using inclusive discriminants.
This yields an estimate of the total four--body decay rate, including
narrow resonant, wide resonant and non resonant final state hadronic systems.

The most recent results from ALEPH~\cite{A_bldst} and 
DELPHI~\cite{D_bldst} are reported in Table~\ref{tab:bldst}.

\begin{table*}
\begin{center}
\caption{Narrow resonant \dst\ production and inclusive four--body 
rate in semileptonic $b$ decays.
\label{tab:bldst}
}
\begin{tabular}{lcc} 
\br
& ALEPH & DELPHI \\
& (published) & (preliminary)\\
\mr
BR($B\to D^{(\star)} \pi \ell \nu (X)$ )
& $0.0214 \pm 0.0041$ 
& $0.0343 \pm 0.0061$ \\
BR($B\to D_1 \ell \nu$)
& $0.0070 \pm 0.0015$ 
& $0.0072 \pm 0.0026$ \\
BR($B\to D^{\star}_2 \ell \nu$) 
& $< 1.5 \cdot 10^{-3} \ @\ 95\% \ {\mathrm{CL}}$ 
& not seen \\
\br
\end{tabular}
\end{center}
\end{table*}

The inclusive four--body decay rate observed by DELPHI would account
for the difference between the total semileptonic rate and the sum of
$D$ and $D^{\star}$ rates. The corresponding ALEPH results is lower by 
$1.8 \sigma$. No significant production of \dstt\ is observed, in contrast
with what is given by all available theoretical predictions.

\boldmath
\section{Production of \dsst\ mesons in $Z$ decays}
\unboldmath
\dsst\ mesons are reconstructed in the channels {\mbox{$D^{\star} K$}}
and {\mbox{$D K$}}. The {\mbox{$D_{s1}^{\star +}$}} is searched for in the
{\mbox{$D^{\star 0} K^+$}} and {\mbox{$D^{\star +} K^0_S$}} final states.
The second has a cleaner experimental signature but lower efficiency due
to the $K^0_S$ reconstruction, 
and lower branching ratio both due to phase space (the $Q$--value of the decay 
is very small) and to the factor $1/2$ because of the  $K^0_S$. 
The $D^{\star +}$ is reconstructed in the  {\mbox{$D^0 \pi^+$}} 
final state, with {\mbox{$D^0 \to K^- \pi^+$}} in the case of 
OPAL~\cite{O_d},
while ALEPH~\cite{A_dsst}
 uses also {\mbox{$D^0 \to K^- \pi^+ \pi^- \pi^+$}} and 
{\mbox{$D^0 \to K^- \pi^+ \pi^0$}}. For the {\mbox{$D^{\star 0} K^+$}}
final state, the $D^{\star 0}$ decays to {\mbox{$D^0 \pi^0$}} or 
{\mbox{$D^0 \gamma$}}, where the photon or the pion are not detected.
The  $D^0$ is reconstructed in the channel  {\mbox{$D^0 \to K^- \pi^+$}},
and the resolution in the 
{\mbox{$D_{s1}^{\star +} - D^0$}} 
mass difference 
only marginally suffers from the missing $\pi^0$ or $\gamma$.
Both ALEPH and OPAL observe resonant structures around 530~MeV/$c^2$.

The  {\mbox{$D_{s2}^+$}} decays to  {\mbox{$D^0 K^+$}}, therefore 
analyzing the {\mbox{$D^0 \to K^- \pi^+$}} channel
both the  {\mbox{$D_{s2}^+$}}  and the  {\mbox{$D_{s1}^{\star +}$}} (with
the missing $\pi^0$ or $\gamma$) can be observed. ALEPH finds a signal
at high mass difference (around 700~MeV/$c^2$), which is interpreted
as the contribution of the  {\mbox{$D_{s2}^+$}}, while for OPAL
resolution and efficiency are estimated to be insufficient to observe
this state.

In the case of ALEPH, kinematic cuts are loose enough to have 
significant reconstruction efficiency also for $\dsst$ from $b$ decays.
The $c$ and $b$ contributions are disentangled in a way similar to that
used in $\dst$ analyses, mentioned in Sect.~\ref{sec:boh}.
The OPAL analysis is sensitive only to $\dsst$ produced in the
hadronization of charm quarks.

The results are summarized in Table~\ref{tab:dsst}.

\begin{table}
\begin{center}
\caption{Results on \dsst\ production probabilities in $Z$ hadronic decays.
\label{tab:dsst}
}
\begin{tabular}{ccc} 
\br
& ALEPH & OPAL \\
& (preliminary) & (published)\\
\mr
$f(c\to D_{s1})$ 
& $0.0077 \pm 0.0022 $& $0.016 \pm  0.005$ \\
$f(c\to D^{\star}_{s2})$ 
& $0.013\ \, \pm 0.019\ \, $& $-$\\
$f(b\to D_{s1})$ 
& $0.011\ \, \pm 0.014\ \, $& $-$ \\
$f(b\to D^{\star}_{s2})$ 
& $0.033\ \, \pm 0.020\ \, $& $-$ \\
\br
\end{tabular}
\end{center}
\end{table}

\boldmath
\section{Production of \dstp\ mesons in $Z$ decays}
\unboldmath

Relativistc quark models predict the masses of the \dpri\ and \dstp\ mesons
to be around 2.6~GeV/$c^2$. Although no firm prediction is given for 
their widths, it is considered likely that these states 
could be narrow. The reconstruction of \dstp\ states has been attempted
by DELPHI~\cite{D_dstpr} and OPAL~\cite{O_dstpr}, in the final state 
{\mbox{$D^{\star +} \pi^+ \pi^-$}}, with
{\mbox{$D^{\star +} \to D^0 \pi^+$}}. The $D^0$ is reconstructed
in the channel 
{\mbox{$D^0 \to K^- \pi^+$}} in the case of OPAL, while DELPHI
uses the channel {\mbox{$D^0 \to K^- \pi^+ \pi^- \pi^+$}} as well.

\begin{figure}
  \begin{center}
         \mbox{\epsfig{figure=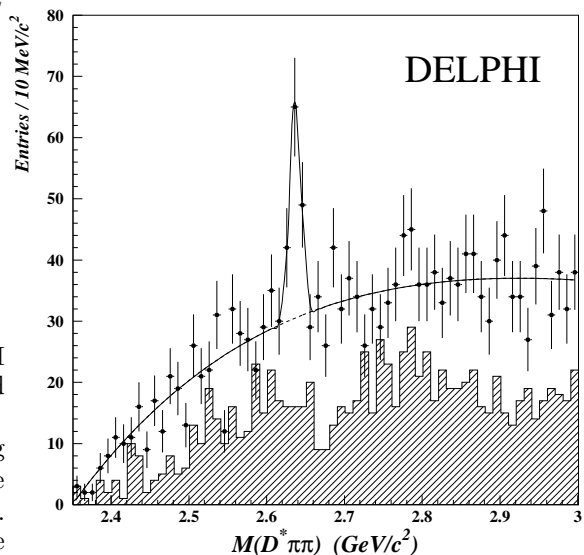,height=75mm,%
              bbllx=2cm,bblly=6.5cm,bburx=20cm,bbury=24cm}} 
   \end{center}
\caption{Reconstructed mass distribution 
of \dstp\ candidates \label{fig:dstpri} }
\end{figure}

A clear excess of events is found by the DELPHI collaboration, with 
a reconstructed mass of
{\mbox{$2637\pm 2 \pm 6$}}~MeV/$c^2$
(see Fig.~\ref{fig:dstpri}). 
The signal events in the peak are 
estimated to be $66\pm 14$ which gives a significance of $4.7\sigma$.
The width of the peak is compatible with the experimental resolution,
therefore no value is given, but the limit $\Gamma < 15$~MeV is set at 95\% CL.

Given the signal observed by DELPHI, OPAL would be expected to see a 
significant signal as well, despite the somewhat worse mass resolution
and the use of the {\mbox{$D^0 \to K^- \pi^+$}} channel only. Instead,
no evidence for a resonance is found. In order to quantify in a conservative
way the level of incompatibility of the two observations, the ratio has been
calculated, between the number of  {\mbox{$D^{\star \prime +}$}} 
candidates reconstructed in the
{\mbox{$D^{\star +} \pi^+ \pi^-$}} channel, and the number of 
{\mbox{$D_1^0, D^{\star 0}_2$}} candidates reconstructed in the 
{\mbox{$D^{\star +} \pi^-$}} final state. In the ratio most of the 
experimental effects cancel out, so that the values obtained by the DELPHI
and the OPAL experiments can be directly compared, although, on the other
hand, the significance of the incompatibility between the two results 
is substantially washed out by the small statistics of the
sample used for the normalization.

The results obtained are $R= 0.49 \pm 0.18 \pm 0.10$ for DELPHI and
$R < 0.21\ @\ 95\%$~CL for OPAL.

\section{Conclusions}
Narrow orbital excitations of charmed and charmed strange 
mesons have been observed at LEP
in $Z$ decays. The contributions of direct $c$ quark hadronization
and $b$ hadron decays have been separated on a statistical basis.
Some differences are observed between the rates measured by different
experiments.

The production rates of \dst\ states and the inclusive four--body rate
have been measured in semileptonic $b$ decays. 

The observation of a charmed meson radial excitation by DELPHI has
not been confirmed by OPAL.

\end{document}